\documentstyle[prd,aps,epsf,eqsecnum]{revtex}

\begin{document}

\title{Time evolution of the chiral phase transition during a spherical
expansion}
\author{Melissa A. Lampert\thanks{e-mail:melissa.lampert@unh.edu}
        and John F. Dawson\thanks{e-mail:john.dawson@unh.edu}}
\address{Department of Physics \\
University of New Hampshire \\
Durham, NH 03824, USA \\ 
}
\author{Fred Cooper\thanks{e-mail:cooper@pion.lanl.gov}}
\address{Theoretical Division \\
Los Alamos National Laboratory \\
Los Alamos, NM 87545, USA \\
}
\preprint{LA-UR-96}
\date{\today}
\maketitle
\begin{abstract}
{We examine the non-equilibrium time evolution of the hadronic
plasma produced in a relativistic heavy ion collision, assuming a
spherical expansion into the vacuum. We study the $O(4)$ linear sigma
model to leading order in a large-$N$ expansion.  Starting at a
temperature above the phase transition, the system expands and cools,
finally settling into the broken symmetry vacuum state. We consider
the proper time evolution of the effective pion mass, the order
parameter $\langle \sigma \rangle$, and the particle number
distribution.  We examine several different initial conditions and
look for instabilities (exponentially growing long wavelength modes)
which can lead to the formation of disoriented chiral condensates
(DCCs). We find that instabilities exist for proper times which are
less than 3 fm/c. We also show that an experimental signature of
domain growth is an increase in the low momentum spectrum of outgoing
pions when compared to an expansion in thermal equilibrium. In
comparison to particle production during a longitudinal expansion, we
find that in a spherical expansion the system reaches the ``out''
regime much faster and more particles get produced. However the size
of the unstable region, which is related to the domain size of DCCs,
is not enhanced.}
\end{abstract}

\section{Introduction}

There have been many recent investigations into the formation of
disoriented chiral condensates (DCCs) following a relativistic heavy-ion
collision\cite{ref:anselm,ref:bjorken,ref:rajwil}. The original
motivation for studying this problem was the Centauro
events\cite{ref:anselm}, rare cosmic ray events in which a deficit of
neutral pions was observed\cite{ref:cent}. In a recent work
\cite{ref:cooperdcc}, the time evolution of the hadronic plasma
produced in such a collision was studied by using the $O(4)$ linear
sigma model in a longitudinal expansion. The large-$N$ expansion was
used to incorporate non-equilibrium and quantum effects into the
problem. After performing numerical simulations to solve the
time-dependent equations of motion, instabilities were found to exist
for only a short time, and thus no significant amount of pion domains
would be formed. In this work, we study the same problem using a
spherical expansion, since at late times the expansion becomes
spherical. This situation produces the most rapid cooling of the
system. We would like to see if the formation of instabilities in this
geometry is more pronounced than in a longitudinal expansion.

There are two questions which should be examined. First, we want to
know which types of initial conditions lead to the formation of
instabilities in the system; and second, if instabilities do form, we
want to find out if the size of the unstable region is large enough
to see significant domain growth. To answer the first question, we
examine the proper-time evolution of the system, starting a short time
after the phase transition, where the linear sigma model is
appropriate. We look for the effective mass of the pion to go negative
during the time evolution. This signifies the onset of growth in
long-wavelength modes, which is believed to lead to the formation of
DCCs. When we have a case where instabilities form, we then compute
the momentum distribution of outgoing pions, and compare to a
hydrodynamical model calculation, assuming local thermal
equilibrium. We find a noticeable enhancement of low momentum modes as
compared to the hydrodynamical model. This provides an experimental
signature which can be measured. The implication is that the system is
evolving out of thermal equilibrium, which is a necessary condition to
have significant growth of low momentum modes.

We prepare the initial state of the system in local thermal
equilibrium, and study the evolution using scale invariant kinematics
($v=r/t$) to model the cooling of the plasma. Scale invariant
kinematics are appropriate for a high energy spherical expansion
starting from a point source
\cite{ref:bjorkenlandau,ref:cooperlandau}, and imply that mean-field
expectation values only depend on the proper time $\tau =
\sqrt{t^2-r^2}$.  We incorporate non-equilibrium and quantum effects
through the use of the large-$N$ expansion.

The paper is organized as follows. In Sec.~\ref{sec:sigmod} we
describe the linear sigma model to leading order in a $1/N$
expansion. We then discuss in Sec.~\ref{sec:coords} our choice of
coordinates and derive equations of motion for the system. We examine
issues of renormalization and choice of suitable initial
conditions. We also derive an expression for the phase space number
density that will be used to calculate momentum distributions. In
Sec.~\ref{sec:tmunu} we discuss the calculation of the energy-momentum
tensor, and other thermodynamic relations. In Sec.~\ref{sec:results}
we present numerical results of our simulation. In Appendix
\ref{app:piprop} we look at properties of the radial functions used in
the expansion for the quantum modes. In Appendix \ref{app:nphys} we
discuss the transformation of the number density to physically
measurable variables.

\section{Linear Sigma Model}
\label{sec:sigmod}

The Lagrangian density for the linear sigma model in a generalized
curvilinear coordinate system is given by:
\begin{eqnarray}
   {\cal L}[\Phi_i] &=& \sqrt{-g(x)} \biggl\{\frac{1}{2} \,
      g_{\mu\nu}(x) [ \partial^\mu \Phi_i(x) ][ \partial^\nu \Phi_i(x)
      ] \nonumber \\ && \phantom{\sqrt{-g(x)} \biggl\{ } -
      \frac{\lambda}{4}\left[\Phi_i^2(x) - v^2 \right]^2 \biggr\} \>,
\end{eqnarray}
where the mesons are in a $O(4)$ vector, $\Phi =
(\sigma,\vec\pi)$. The factor of $\sqrt{-g(x)}$, where $-g(x) \equiv
{\rm det}[g_{\mu\nu}(x)]$, has been introduced to make the Lagrangian
a scalar density.  The potential here is the ``Mexican hat'', with
degenerate minima at any values of $\Phi$ such that $\Phi_i^2 = v^2$.
We will remove this symmetry by introducing a non-zero current term in
the $\sigma$ direction.  In this work, we use the convention of an
implied sum over a repeated index $i$, which runs from 1 to $N$. (Here
$N = 4$.) The counting for the large-$N$ expansion is implemented by
introducing a composite field $\chi = \lambda ( \Phi_i^2 - v^2 )$.
That is, we add to the Lagrangian a term given by\cite{ref:Stanley}:
\begin{displaymath}
   \left[ \, \chi - \lambda ( \Phi_i^2(x) - v^2 ) \, \right]^2 / \, 4
      \lambda \>.
\end{displaymath}
This gives an equivalent Lagrangian,
\begin{eqnarray}
   {\cal L}[\Phi_i, \chi] &=& \sqrt{-g(x)} \biggl\{ \frac{1}{2}
      g_{\mu\nu}(x) [ \partial^\mu \Phi_i ][ \partial^\nu \Phi_i ]
      \nonumber \\ && \phantom{\sqrt{-g(x)} \biggl\{ } - \frac{1}{2}
      \chi\Phi_i^2 + \frac{v^2}{2} \chi + \frac{1}{4 \lambda} \chi^2
      \biggr\} \>.
\label{eq:lag}
\end{eqnarray}

We can write the action as:
\begin{equation}
   S[\Phi_i, \chi; j_i] = \int d^4 x \, \left\{ {\cal L}[\Phi_i, \chi]
      + \sqrt{-g(x)} j_i \Phi_i \right\} \>.
\label{eq:action}
\end{equation}

We consider the generating functional, given by the path integral,
\begin{displaymath}
   Z[j_i] = \int {\rm d}[\chi] \int {\rm d}[\Phi_i] \, \exp \bigl[i
      S[\Phi_i, \chi; j_i]\bigr] \equiv \exp \bigl[i W[j_i] \bigr] \>.
\end{displaymath}

The large-$N$ approximation is equivalent to integrating out the
$\Phi_i$ variables, then performing the remaining $\chi$ integral
using the method of steepest descent. We then Legendre transform to
find the effective action,

\begin{displaymath}
   \Gamma[\phi_i, \chi] = W[j_i] - \int d^4 x \, \sqrt{-g(x)} j_i(x)
      \phi_i(x) \>,
\end{displaymath}
where
\begin{displaymath}
   \phi_i(x) = \frac{\delta W}{\delta j_i} \equiv \langle \Phi_i(x)
      \rangle \>.
\end{displaymath}

We then obtain, to lowest order (in large N)
\begin{eqnarray}
   \Gamma[\phi_i,\chi] &=& \int d^4 x \, \sqrt{-g(x)} \biggl\{
      \frac{1}{2} g_{\mu\nu}(x) [ \partial^\mu \phi_i][ \partial^\nu
      \phi_i ] \nonumber \\ && \phantom{ \int d^4 x \, \sqrt{-g(x)}
      \biggl\{ } - \frac{1}{2} \chi\phi_i^2 + \frac{v^2}{2} \chi +
      \frac{1}{4 \lambda} \chi^2 \nonumber \\ && \phantom{ \int d^4 x
      \, \sqrt{-g(x)} \biggl\{ } + \frac{i N}{2} \ln
      G_0^{-1}(x,x;\chi) \biggr\} \>, \nonumber \\
\label{eq:effaction}
\end{eqnarray}
where
\begin{displaymath}
   (\Box + \chi(x))G_0(x,x';\chi) = i \delta^4(x-x') / \sqrt{-g(x)}
\>.
\end{displaymath}
$\Gamma[\phi_i, \chi]$ is the {\em classical} action plus the
trace-log term.

\section{Cooling Mechanism}
\label{sec:coords}

\subsection{Coordinate System}
At late times following a heavy ion collision, the energy flow becomes
three-dimensional. If we assume the entire flow is spherical, then the
system can be described in terms of the fluid variables:
\begin{eqnarray*}
   \tau & = & \sqrt{ t^2 - r^2 } \>, \\ \eta & = & \displaystyle
   \tanh^{-1} ( r / t ) = \frac{1}{2} \ln \left\{ \frac{ t + r }{ t -
   r } \right\} \>,
\end{eqnarray*}
where $t = \tau\cosh\eta$ and $r = \tau\sinh\eta$.  We restrict the
range of these variables to the forward light cone, $0 \leq \tau <
\infty$, and $0 \leq \eta < \infty$.  The variables $\tau$ and $\eta$
are useful to describe a free spherical expansion of a plasma into the
vacuum, with the velocity of the fluid identified as $v = r/t =
\tanh\eta$ \cite{ref:cooperlandau} when we are at high energies so
that the expansion can be thought of as coming from a point source
(scaling limit).  Minkowski's line element is given by:
\begin{displaymath}
   d s^2 = d \tau^2 - \tau^2 \left\{ d \eta^2 + \sinh^2 \eta \, d
      \theta^2 + \sinh^2 \eta \sin^2 \theta \, d \phi^2 \right\} \>,
\end{displaymath}
from which we can read off the metric tensor
\begin{eqnarray*}
   g_{\mu\nu} &=& \mbox{diag} (1, -\tau^2, -\tau^2 \sinh^2 \eta,
      -\tau^2 \sinh^2 \eta \sin^2 \theta) \>, \\ \sqrt{-g} & = &\tau^3
      \sinh^2 \eta \sin \theta \>.
\end{eqnarray*}
This can be compared to the Robertson-Walker metric for spherical
geometry, given by the line element:
\begin{eqnarray*}
   d s^2 &=& d \tau^2 \\ &-& a^2(\tau) \left\{ d \eta^2 + \sinh^2 \eta
   \, d \theta^2 + \sinh^2 \eta \sin^2 \theta \, d \phi^2 \right\} \>.
\end{eqnarray*}
Thus the case we consider here corresponds to a cosmological model
with a fixed uniform expansion proportional to the proper time $\tau$,
and zero curvature.

\subsection{Equations of motion}

We can derive the equations of motion from the effective action,
(\ref{eq:effaction}).  Varying the action with respect to $\phi_i$ and
$\chi$ gives:
\begin{eqnarray}
   ( \Box + \chi(x) ) \phi_i(x) & = & j_i(x) \equiv H \delta_{i0}
   \nonumber \\ ( \Box + \chi(x) ) G_0(x,x') & = & i \delta^4(x - x')
   / \sqrt{-g(x)} \>,
\label{eq:eom}
\end{eqnarray}
and the gap equation
\begin{equation}
   \chi(x) = \lambda \left\{ -v^2 + \phi_i^2(x) + N G_0(x,x) \right\}
\label{eq:chi}
\>.
\end{equation}
In order to give the pions mass, it is only necessary to have a
current in the zero ($\sigma$) direction, so that $j_0(x) \equiv H =$
constant.

We determine the parameters in the model by considering the vacuum
sector. Spatial homogeneity then gives:
\begin{eqnarray*}
   \chi_0\sigma_0 &=& H \>, \\ \chi_0 &=& -\lambda v^2 + \lambda
   \sigma_0^2 + \lambda N \int \, \frac{k^2 d k}{2\pi^2}
   \frac{1}{2\sqrt{k^2 + \chi_0}} \>.
\end{eqnarray*}
The PCAC (partial conservation of axial vector current) condition
gives:
\begin{displaymath}
   \partial_{\mu} A^{\mu}_i(x) = H \pi_i(x) \>,
\end{displaymath}
where the axial vector current is given by
\begin{displaymath}
   A^{\mu}_i(x) = [\pi_i(x) \partial^{\mu}\sigma(x) - \sigma(x)
      \partial^{\mu} \pi_i(x)] \>,
\end{displaymath}
which leads to
\begin{displaymath}
   H = f_{\pi} m^2 \>.
\end{displaymath}
Since we have defined the vacuum by $\chi_0\sigma_0 = m^2 \sigma_0 =
H$, we then find that $\sigma_0 = f_\pi$. The coupling constant
$\lambda$ is determined by the low-energy $\pi-\pi$ scattering data,
as described in\cite{ref:cooperdcc}.

We now specialize to the case when $\phi_i$ and $\chi$ are functions
of $\tau$ only. We can see that (\ref{eq:eom}) is also the equation
for a free scalar quantum field with a time-dependent mass
$\chi(\tau)$, which is self-consistently determined by
(\ref{eq:chi}). Therefore we can introduce a quantum field $\Phi_i =
\phi_i + \hat\phi_i$.  The equations for $\Phi_i$ are:
\begin{eqnarray}
   \left\{ \frac{1}{\tau^3} \frac{ \partial }{ \partial \tau } \left(
      \tau^3 \frac{ \partial }{ \partial \tau } \right) + \chi(\tau)
      \right\} \phi_i(\tau) &=& H \delta_{i0} \nonumber \\ (\Box +
      \chi(\tau))\hat\phi_i(x) &=& 0 \>,
\end{eqnarray}
where the 4-vector $x = (\tau,\eta,\theta,\phi)$.  Then for $G_0$ we
find:
\begin{displaymath}
   G_0(x,x') \equiv \langle{\rm T}_c\{ \hat{\phi}(x,\tau), \,
      \hat{\phi}(x',\tau')\} \rangle \>,
\end{displaymath}
where ${\rm T}_c$ corresponds to a $\tau$-ordered
product\cite{ref:ctp}, following the closed-time-path formalism of
Schwinger.  When $\langle \pi^i \rangle = 0$, this is the true Green's
function.

Following Parker and Fulling\cite{ref:parker}, we expand $\hat\phi_i$
into a complete set of states,
\begin{eqnarray}
   \hat{\phi}_i(\tau,\eta,\theta,\phi) &=& \int_0^{\infty} d s
      \sum_{lm} \bigl[ \hat{a}_{i,slm} \, \psi_s(\tau)
      {\cal{Y}}_{slm}(\eta,\theta,\phi) \nonumber \\ &&
      \phantom{\int_0^{\infty} d s \sum_{lm} \bigl[ } + {\rm h.c.} \,
      \bigr]
\label{eq:hatphi}
\end{eqnarray}
with
\begin{equation}
   {\cal{Y}}_{slm}(\eta,\theta,\phi) = \pi_{sl}(\eta)
   Y_{lm}(\theta,\phi) \>, \label{eq:bigy}
\end{equation}
and where $\psi$, $\pi$, and $Y$ satisfy:
\begin{displaymath}
   \biggl[\frac{1}{\tau^3} \frac{\partial}{\partial\tau} \biggl(\tau^3
      \frac{\partial}{\partial \tau}\biggr) + \frac{s^2 + 1}{\tau^2} +
      \chi(\tau) \biggr]\psi_s = 0 \>,
\end{displaymath}
\begin{displaymath}
   \biggl[\frac{1}{ \sinh^2 \eta } \frac{\partial}{\partial \eta}
      \biggl( \sinh^2 \eta \, \frac{\partial}{\partial \eta} \biggr) +
      s^2 + 1 - \frac{ l(l+1) }{ \sinh^2 \eta } \biggr]\pi_{sl} = 0
      \>,
\end{displaymath}
\begin{displaymath}
   \biggl[\frac{1}{\sin \theta} \frac{\partial}{\partial \theta}
      \biggl( \sin \theta \, \frac{\partial}{\partial \theta} \biggr)
      + \frac{1}{\sin^2 \theta} \biggl( \frac{\partial^2}{\partial
      \phi^2} \biggr) + l(l+1) \biggr]Y_{lm} = 0 \>.
\end{displaymath}
Here, $Y_{lm}(\theta,\phi)$ are the usual spherical harmonics, and
$\pi_{sl}(\eta)$ are a complete set of functions, discussed in
Appendix \ref{app:piprop}.

In order to satisfy the canonical commutation relations, we require
$\psi_s(\tau)$ to satisfy the Wronskian condition:
\begin{displaymath}
   \psi_s^{\ast}(\tau) \dot{\psi}_s(\tau) - \psi_s(\tau)
      \dot{\psi}_s^{\ast}(\tau) = - i / \tau^3 \>,
\end{displaymath}
where the ``dot'' means differentiation with respect to $\tau$.
Therefore we find
\begin{displaymath}
   [ \hat{\phi}_{i,slm}(\tau),
     \hat{\dot{\phi}}_{j,s'l'm'}^{\dagger}(\tau) ] = i \delta_{ij} \,
     \delta(s' - s) \, \delta_{l l'} \delta_{m m'} / \tau^3 \>,
\end{displaymath}
and
\begin{displaymath}
   [ \hat{a}_{i,slm}, \hat{a}_{j,s'l'm'}^{\dagger} ] = \delta_{i j} \,
      \delta(s' - s) \, \delta_{l l'} \delta_{m m'} \>.
\end{displaymath}
We are now in a position to calculate $\langle \hat\phi_i^2 \rangle$.
We shall choose the (Heisenberg) state vector such that the bilinear
forms of creation and destruction operators are diagonal:
\begin{eqnarray*}
   \langle \hat{a}_{j,s'l'm'}^{\dagger} \hat{a}_{i,slm} \rangle & = &
      n_s \, \delta_{i j} \, \delta(s' - s) \, \delta_{l l'} \delta_{m
      m'} \>, \\ \langle \hat{a}_{j,s'l'm'} \hat{a}_{i,slm}^{\dagger}
      \rangle & = & ( n_s + 1 ) \, \delta_{i j} \, \delta(s' - s) \,
      \delta_{l l'} \delta_{m m'} \>, \\ \langle
      \hat{a}_{j,s'l'm'}\hat{a}_{i,slm} \rangle & = & p_s \, \delta_{i
      j} \, \delta(s' - s) \, \delta_{l l'} \delta_{m m'} \>, \\
      \langle \hat{a}_{j,s'l'm'}^{\dagger}\hat{a}_{i,slm}^{\dagger}
      \rangle & = & p_s^{\ast} \, \delta_{i j} \, \delta(s' - s) \,
      \delta_{l l'} \delta_{m m'} \>.
\end{eqnarray*}
Here $n_s$ and $p_s$ are the particle and pair densities.  They will
be taken to be a function of $s$ only.  In addition, we will take
$n_s$ to be a thermal distribution in the comoving frame,
\begin{displaymath}
   n_s = \frac{1}{e^{\omega_s(\tau_0) / k_B T} - 1} \>,
\end{displaymath}
with $\omega_s = \sqrt{s^2/\tau_0^2 + \chi(\tau_0)}$. We can choose
$p_s = 0$ for all our simulations, since one has the freedom to make a
Bogoliubov transformation at $\tau_0$ so that this is true. Using the
results in Appendix \ref{app:piprop}, we then find
\begin{eqnarray*}
   \langle\hat\phi_i^2\rangle &=& \int_0^{\infty} d s \, ( 2 n_s + 1 )
      | \psi_s(\tau) |^2 \, \sum_{lm} | {\cal
      Y}_{slm}(\eta,\theta,\phi) |^2 \\ &=& \int_0^{\infty} \frac{s^2
      d s}{2\pi^2} \, (2 n_s + 1) | \psi_s(\tau)|^2 \>.
\end{eqnarray*}
Therefore (\ref{eq:chi}) becomes:
\begin{equation}
   \chi(\tau) = - \lambda v^2 + \lambda \phi_i^2(\tau) + \lambda N
      \int_0^{\infty} \frac { s^2 d s }{ 2 \pi^2 } \, ( 2 n_s + 1 ) |
      \psi_s(\tau) |^2 \>,
\label{eq:chig} 
\end{equation}
and is a function of $\tau$ only.  This completes the derivation of
the equations of motion.

\subsection{Initial Conditions and Renormalization}

The variable $\tau$ does not allow for a good WKB expansion due to the
singularity at $\tau=0$. The transformation $u = \ln(m\tau)$, where
$m$ is any mass scale (we choose $m = m_\pi$), maps the singularity to
$u = -\infty$, and allows one to perform a WKB expansion in the usual
manner.

Changing variables to $u$ and rescaling the fields using the
substitutions
\begin{eqnarray*}
   \psi_s(u) &=& g_s(u) e^{-u} m^{3/2} \\ \phi_i(u) &=& \rho_i(u)
   e^{-u} m^{3/2} \>,
\end{eqnarray*}
we get:
\begin{eqnarray}
   \left[\frac{d^2}{du^2} + s^2 + \chi(u) e^{2u}/m^2 \right] g_s(u)
      &=& 0 \nonumber \\ \left[\frac{d^2}{du^2} + \chi(u) e^{2u} / m^2
      - 1 \right] \rho_i(u) &=& H \delta_{i0} e^{3u} / m^{7/2}
\label{eq:ueom}
\end{eqnarray}
with
\begin{eqnarray}
   \chi(u) &=& - \lambda v^2 + \lambda \sum_i m^3 e^{-2u} \rho_i^2(u)
   \nonumber \\ &+& \lambda N \int_0^{s_m} \frac{ s^2 d s}{ 2 \pi^2 }
   \, (2 n_s + 1) m^3 e^{-2u} | g_s(u)|^2 \>,
\label{eq:chiu}
\end{eqnarray}
with $s_m = \Lambda e^u / m$.
Then $g_s(u)$ obeys the Wronskian
\begin{displaymath}
   g_s^{\ast}(u) \frac{dg_s(u)}{du} - \frac{dg_s^{\ast}(u)}{du} g_s(u)
   = - i/m \>.
\end{displaymath}
From (\ref{eq:ueom}), we notice that when $\chi(u) < 0$, modes satisfying 
$s^2 / \tau^2 < \chi$ grow exponentially in $\tau$. However, from 
(\ref{eq:chiu}), we see that these growing modes will quickly cause 
$\chi$ to become positive when $\lambda$ is large, as is the case here.

We can now use a WKB ansatz for $g_s(u)$:
\begin{displaymath}
   g_s(u) = \frac{1}{\sqrt{2 m W_s(u)}} \, {\rm exp} \left[
      -i \int_{u_0}^u \, W_s(u') du' \right] \>,
\end{displaymath}
where $W_s(u)$ satisfies:
\begin{equation}
   \frac{1}{2} \frac{W_s{''}}{W_s} - \frac{3}{4}
      \left( \frac{W'_s}{W_s} \right)^2 + W_s^2 = \omega_s^2(u)
\>,
\end{equation}
and $\omega_s(u) = \sqrt{s^2 + \chi(u) e^{2u} / m^2}$.
We will then take the initial conditions 
\begin{eqnarray*}
   W_s(u_0) &=& \omega_s(u_0)
   \\
   W'_s(u_0) &=& \omega'_s(u_0)
\>,
\end{eqnarray*}
which correspond to the adiabatic vacuum. This allows us to introduce
an interpolating number density which interpolates between $n_s(u)$
and the true ``out'' density, $n_{{\rm out}}$. 

By a WKB analysis, one can show\cite{ref:cooperren} that $G_0(x,x)$ has
quadratic and logarithmic divergences. The quadratic divergence can be
removed by mass renormalization. In the vacuum sector, the mass of the
pion
is given by (\ref{eq:chiu}), with $\chi_{\rm{vac}} \equiv m_\pi^2 =
m^2$:
\begin{equation}
   m^2 = - \lambda v^2 + 
      \lambda f_\pi^2 + 
      \lambda N \int_0^{\Lambda}\frac{k^2 dk}{ 2 \pi^2 }
      \, \frac{1}{2 \sqrt{ k^2 + m^2 }}
\>,
\label{eq:vacmass}
\end{equation}
with cutoff $\Lambda$.  
We note that if we change variables in the integral to 
$s = k \tau = k e^u/m$, (\ref{eq:vacmass}) becomes: 
\begin{eqnarray}
   m^2 = -&& \lambda v^2 + 
      \lambda f_\pi^2 
   \nonumber \\
   +&& \lambda N m^2 e^{-2u} 
      \int_0^{s_m}
      \frac{s^2 ds}{ 2 \pi^2 }
      \, \frac{1}{2 \sqrt{ s^2 + e^{2u} }}
\>.
\label{eq:massu}
\end{eqnarray}
Subtracting this equation from (\ref{eq:chiu}), we obtain a
logarithmically divergent expression for $\chi$:
\begin{eqnarray}
   \chi(u)& = & m^2 + \lambda \sum_i 
      \left( m^3 e^{-2u} \rho_i^2(u) - f_{\pi}^2 \right)
   \nonumber \\
   &&
   + \lambda N \int_0^{s_m} 
      \frac{ s^2 ds}{ 2 \pi^2 } \,
      (2 n_s + 1) \, m^3 e^{-2u} | g_s(u)|^2
   \nonumber \\
   && - \lambda N \int_0^{s_m} 
      \frac{ s^2 ds}{ 2 \pi^2 } \,
      m^2 e^{-2u} \frac{1}{2 \sqrt{s^2 + e^{2u}}}
\>.
\label{eq:chiren}
\end{eqnarray}
Note that the last integral is {\em independent} of $u$.

The coupling constant is renormalized by taking
\begin{eqnarray}
   \frac{1}{\lambda} 
      & = &  \frac{1}{\lambda_r} -  
      \frac{N}{8 \pi^2} \int_0^{\Lambda} 
      \frac { k^2 dk }{ ( k^2 + m^2 )^{3/2} }
   \nonumber \\
   & = &  \frac{1}{\lambda_r} -  
      \frac{N}{8 \pi^2} \int_0^{s_m} 
      \frac { s^2 ds }{ ( s^2 + e^{2u} )^{3/2} }
\>.
\label{eq:lamren}
\end{eqnarray}
One can explicitly show by using (\ref{eq:lamren}) in
(\ref{eq:chiren}) that $\chi(u)$ is now completely finite.

We will also need the value of $\dot\chi(u)$ for the initial
conditions:
\begin{eqnarray}
   && \dot\chi(u) \left[ 1 + \lambda N \int_0^{s_m} \, 
      \frac{s^2 ds}{2\pi^2} (2 n_s + 1) 
      \frac{1}{4 \omega_s^3(u)}\right]
   \nonumber \\
   &=& 2 \lambda e^{-2u}m^3 [\rho_i(u) \rho'_i(u) 
      - \rho_i^2(u)] 
   \nonumber \\
   &+& \frac{\lambda N \Lambda^3}{4\pi^2}
      \frac{(2 n_{s_m} + 1)}
      {\sqrt{\Lambda^2 + \chi(u)}}
   \nonumber \\
   &-&\lambda N \int_0^{s_m} \, 
      \frac{s^2 ds}{2\pi^2} (2 n_s + 1)
      m^2 e^{-2u}
   \nonumber \\
   && \phantom{\lambda N \int_0^{s_m}} 
      \times \left[ \frac{\chi(u)e^{2u}}
      {2 m^2 \omega_s^3(u)} + \frac{1}{\omega_s(u)}\right]
\>.
\end{eqnarray}

\subsection{Phase Space Interpolating Number Density}

In the expansion (\ref{eq:hatphi}), we can also use a
time-dependent set of creation and annhilation operators, with first
order adiabatic mode functions,
\begin{eqnarray}
   \hat{\phi}_i(u,\eta,\theta,\phi) &=& 
      \int_0^{\infty} d s
      \sum_{lm} \bigl[ \hat{a}_{i,slm}(u) \,
      \psi_s^0(u) {\cal{Y}}_{slm}(\eta,\theta,\phi)
   \nonumber \\
   && \phantom{  \int_0^{\infty} d s
      \sum_{lm} \bigl[ }
      + {\rm h.c.} \, \bigr]
\end{eqnarray}
where
\begin{displaymath}
   \psi_s^0(u) \, = g_s^0(u) e^{-u} m^{3/2}
\>,
\end{displaymath}
and
\begin{displaymath}
   g_s^0(u) = \frac{1}{\sqrt{2 m \omega_s(u)}} 
      \, {\rm exp} \left[-i \int_{u_0}^{u} \, 
      \omega_s(u') du' \right]
\>.
\end{displaymath}
We can define the first order adiabatic number density as:
\begin{equation}
   n_s(u) = \langle \hat{a}^{\dag}_s(u) \hat{a}_s(u) \rangle
\>,
\end{equation}
where, for simplicity, we have suppressed the angular indices on
$\hat{a}_s$ and $\hat{a}^{\dag}_s$.

One can show\cite{ref:mottola} that $n_s(u)$ is an adiabatic
invariant, and would be the true number density in a slowly varying
expansion. We then choose
\begin{eqnarray*}
   \hat{a}_s &=& \hat{a}_s(u_0)
   \\
   g_s(u_0) &=& g_s^0(u_0)
\end{eqnarray*}
so that the initial $\hat{a}$ and $\hat{a}^{\dag}$ are the adiabatic
ones. When $\chi(u) \rightarrow m^2$ then $n_s(u) \rightarrow n_{{\rm
out}}$, which is the true out-state phase space number density.

The time-dependent creation and annihilation operators satisfy
\begin{equation}
   \frac{d \hat{a}_s}{du} g_s^0 
      + \frac{d\hat{a}^{\dag}_s}{du} g^{0 \, \ast} = 0
\>.
\end{equation}

The time-dependent operators can be connected to the time-independent
operators via a Bogoliubov transformation:
\begin{equation}
   \hat{a}_s(u) = \alpha(s,u) \hat{a}_s + \beta(s,u) \hat{a}^{\dag}_{-s}
\>,
\end{equation}
where $\alpha$ and $\beta$ are determined by:
\begin{eqnarray}
   \alpha(s,u) &=& i \left( g_s^{0 \, \ast} 
      \frac{dg_s}{du} 
      - \frac{d g_s^{0 \, \ast}}{du} 
      g_s \right)
   \nonumber \\
   \beta(s,u) &=& i \left( g_s^0 
      \frac{d g_s}{du} 
      - \frac{dg_s^0}{du} 
      g_s \right) \>.
\end{eqnarray}
We then find
\begin{equation}
   n_s(u) = n_s(u_0) + |\beta(s,u)|^2 (1 + 2 n_s(u_0))
\label{nstau}
\>.
\end{equation}
Notice that at $u = u_0$, $\beta(s, u_0) = 0$, so $n_s(u)
= n_s(u_0) \equiv n_s$. Since $n_s(u_0)$ is the initial phase space
number density, and at late times, becomes the out-state number
density, it is an interpolating number density.

\section{Energy-Momentum Tensor}
\label{sec:tmunu}

The energy-momentum tensor $T^{\mu\nu}$ is defined by:
\begin{equation}
   \delta S = - \frac{1}{2} \int d^4 x \sqrt{-g} 
      \, T^{\mu\nu}(x) \, \delta g_{\mu\nu}(x)
\>,
\end{equation}
with the action given by (\ref{eq:action}).  Performing the
variations, we find the ``improved'' energy-momentum
tensor\cite{ref:callan}
\begin{eqnarray}
   T_{\mu\nu} (x) &=& ( \partial_{\mu} \Phi_i ) \, ( \partial_{\nu}
      \Phi_i ) - g_{\mu\nu} {\cal L}
   \nonumber \\
   &&\phantom{(} 
      + \frac{1}{6} \left[g_{\mu\nu}
      g^{\alpha\beta}\Phi_{i;\alpha;\beta}^2 - \Phi_{i;\mu;\nu}^2
\right]
\>,
\end{eqnarray}
where the Lagrangian density is given by (\ref{eq:lag}). We 
follow the standard practice\cite{ref:cooperlandau} and define the
energy density and pressures by
\begin{displaymath}
   T_{\mu\nu} = {\rm diag} \, ( \epsilon, p_\eta \, \tau^2, 
      p_\theta \, \tau^2 \sinh^2 \eta , 
      p_\phi \, \tau^2 \sinh^2 \eta \sin^2 \theta )
\>.
\end{displaymath}
 
Next, we take expectation values of the energy-momentum tensor.
One can easily show all the pressures are equal, $p = p_\eta =
p_\theta = p_\phi$, and that $T_{\mu\nu}$ is diagonal. 

The energy-momentum tensor obeys a conservation law:
\begin{equation}
   T^{\mu\nu}_{;\mu} = \frac{1}{\sqrt{-g}} 
      \frac{\partial}{\partial x^\mu} 
      \left( \sqrt{-g} \, T^{\mu\nu} \right) +
      \Gamma^{\nu}_{\mu\lambda} \, T^{\mu\lambda} = 0
   \>.
\label{tmunu}
\end{equation}
One can easily verify that the energy conservation equation, the $\nu
= 0$ component of (\ref{tmunu}) takes the form:
\begin{equation}
   \frac{ \partial \epsilon }{ \partial \tau } + 
      \frac{3}{\tau}(\epsilon + p)  = 0  \>.
\end{equation}
In thermal equilibrium, this would become the entropy conservation
equation.  Then we have $d\epsilon = T ds$ and $\epsilon + p = T s$.
In an ultra-relativistic fluid expansion, with $c_0^2 = \frac{1}{3}$,
$p = \frac{1}{3} \epsilon$ and $\epsilon \tau^4$ = constant.

In our simulations we have verified that the local energy
conservation (\ref{tmunu}) is valid.

\section{Numerical Results}
\label{sec:results}

To choose the initial conditions, we start the system at
a temperature above the phase transition in thermal equilibrium, with
all particle masses positive. The equations are solved
self-consistently at the starting time to obtain the values of the
$\chi$, $\langle\sigma\rangle$ and $\langle\vec{\pi}\rangle$
fields. We fixed the value of $\chi$ at the initial time as the
solution of the gap equation in the initial thermal state. We also
required that the initial expectation values of the $\sigma$ and
$\vec\pi$ fields satisfy
\begin{displaymath}
   \pi^2(\tau_0) + \sigma^2(\tau_0) = \sigma^2_T
\>,
\end{displaymath}
where $\sigma_T$ is the equilibrium value of $\Phi$ at the initial
temperature $T$.  We choose $T =$ 200 MeV, which gives $\sigma_T =
0.3$ fm $^{-1}$.  We compute the time evolution of these fields,
starting at a proper time $\tau_0 = 1 \, {\rm fm}$. The value of
$f_{\pi}$ used in all the simulations is 92.5 MeV, and $\lambda_R$ is
7.3 (see\cite{ref:cooperdcc}).  Below we show results for several sets
of initial conditions. Once the initial values are chosen, we have the
freedom to vary the first derivative of the $\Phi$ field. The results
with $\dot\pi \ne 0$ were similar to those with $\dot\sigma \ne 0$, so
we only show results for the latter case.  We find there is a wide
range of values which will allow the system to become unstable, $0.15
< |\dot\sigma| < 4.95$. This can be compared with the longitudinal
expansion \cite{ref:cooperdcc}, where the regime of instability was
much smaller, $0.25 < |\dot\sigma| < 1.3$. This is because the
spherical expansion leads to a much larger negative gradient for
$\chi$ than the longitudinal case.

\begin{figure}
\epsfxsize = 6.0in
\centerline{\epsfbox{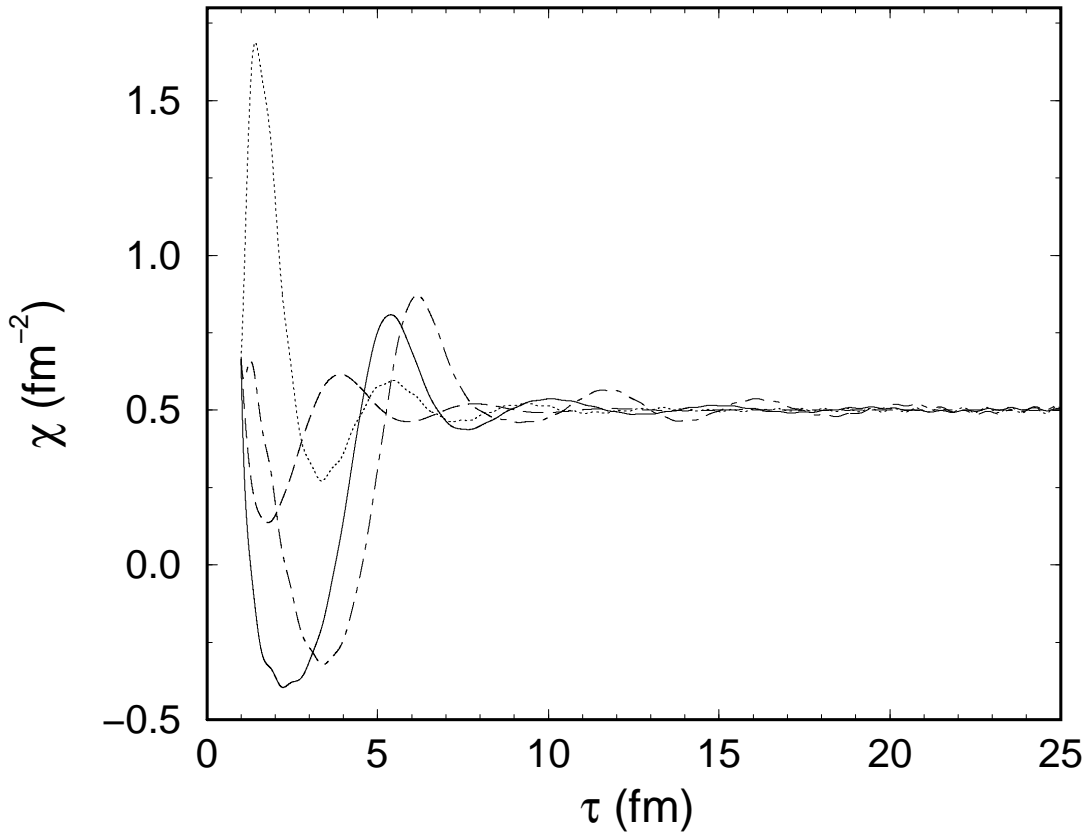}}
\caption{Proper time evolution of the $\chi$ field for the following
initial conditions: Solid line is for $\sigma(\tau_0) = \sigma_{T}$,
$\pi_i(\tau_0) = 0$, and $\dot\sigma(\tau_0) = -1$. Dashed line is for
$\sigma(\tau_0) = \sigma_{T}$, $\pi_i(\tau_0) = 0$, and
$\dot\sigma(\tau_0) = 1$. Dotted line is for $\sigma(\tau_0) =
\sigma_{T}$, $\pi_i(\tau_0) = 0$, and $\dot\sigma(\tau_0) =
0$. Dot-dash line is for $\sigma(\tau_0) = 0$, $\pi_1(\tau_0) =
\sigma_{T}$, and $\dot\sigma(\tau_0) = -1$. At $T = 200$ MeV,
$\sigma_{T} = 0.3 \, {\rm fm}^{-1}$.}
\label{fig:chi}
\end{figure}

\begin{figure}
\epsfxsize = 6.0in
\centerline{\epsfbox{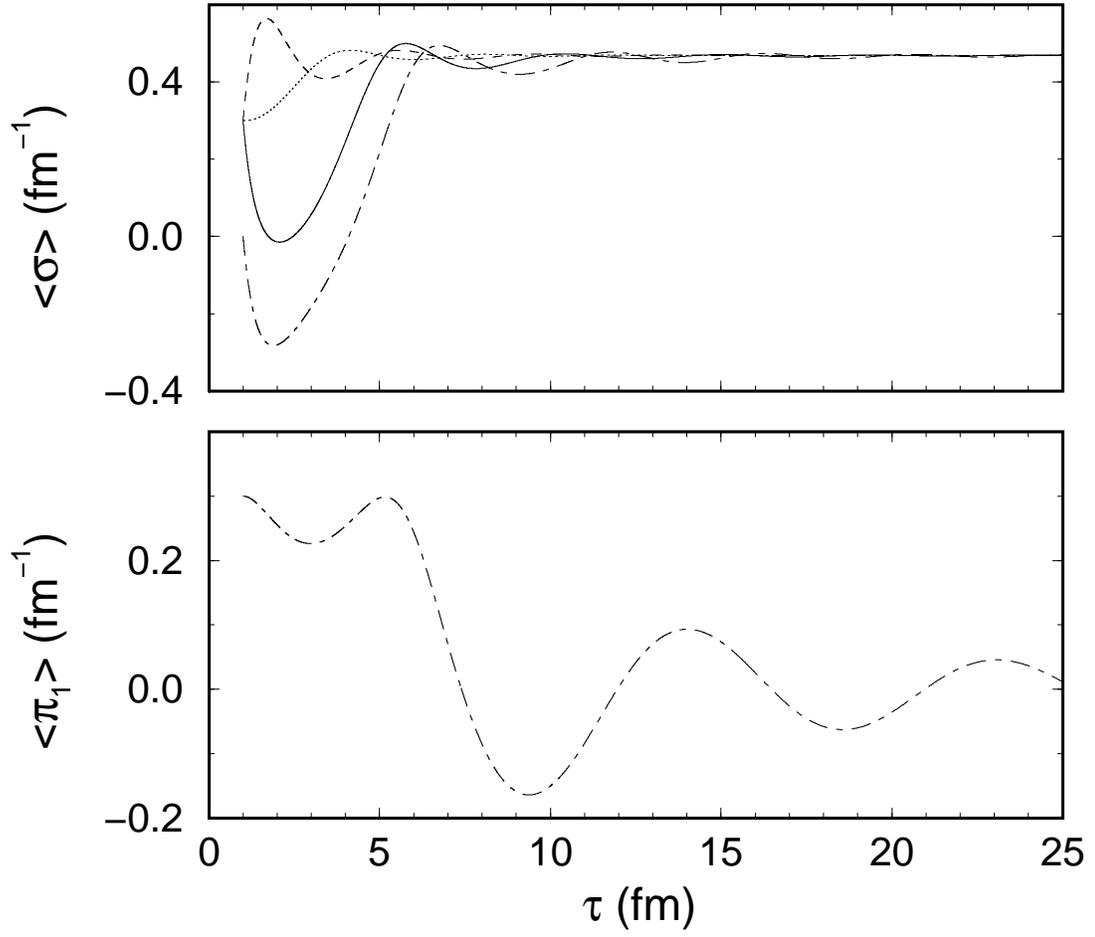}}
\caption{Proper time evolution of the $\langle\sigma\rangle$ and
$\langle\pi_1\rangle$ fields for the same initial conditions as
Fig.~\ref{fig:chi}.}
\label{fig:sigpi}
\end{figure}

Figures \ref{fig:chi} and \ref{fig:sigpi} show the results of the
numerical simulation for the proper time evolution of the system. We
display the auxiliary field $\chi$ in units of fm$^{-2}$, the
classical fields $\Phi_i$ in units of fm$^{-1}$, and the proper time
in units of fm. At proper times greater than $\approx 10$ fm, the
auxiliary field reaches its vacuum value of $m_{\pi}^2$, the $\sigma$
field approaches its vacuum value of $f_{\pi}$, and the $\pi$
field its value of zero. This is in distinction to the longitudinal
expansion, where even at $\tau = 30$ fm, one had not yet reached the
``out'' regime.

For all of our initial conditions, the size of the unstable region
($\chi < 0$) is at most 2-3 fm. Thus we find that the spherical
expansion does not produce larger domains than the longitudinal case.
We also see that when we start the initial expectation value of $\Phi_i$
in the $\pi_1$ direction that the system becomes slightly more
unstable. We notice that if the derivative of the field is positive or
zero that this is insufficient to generate instabilities.

\begin{figure}
\epsfxsize = 6.0in
\centerline{\epsfbox{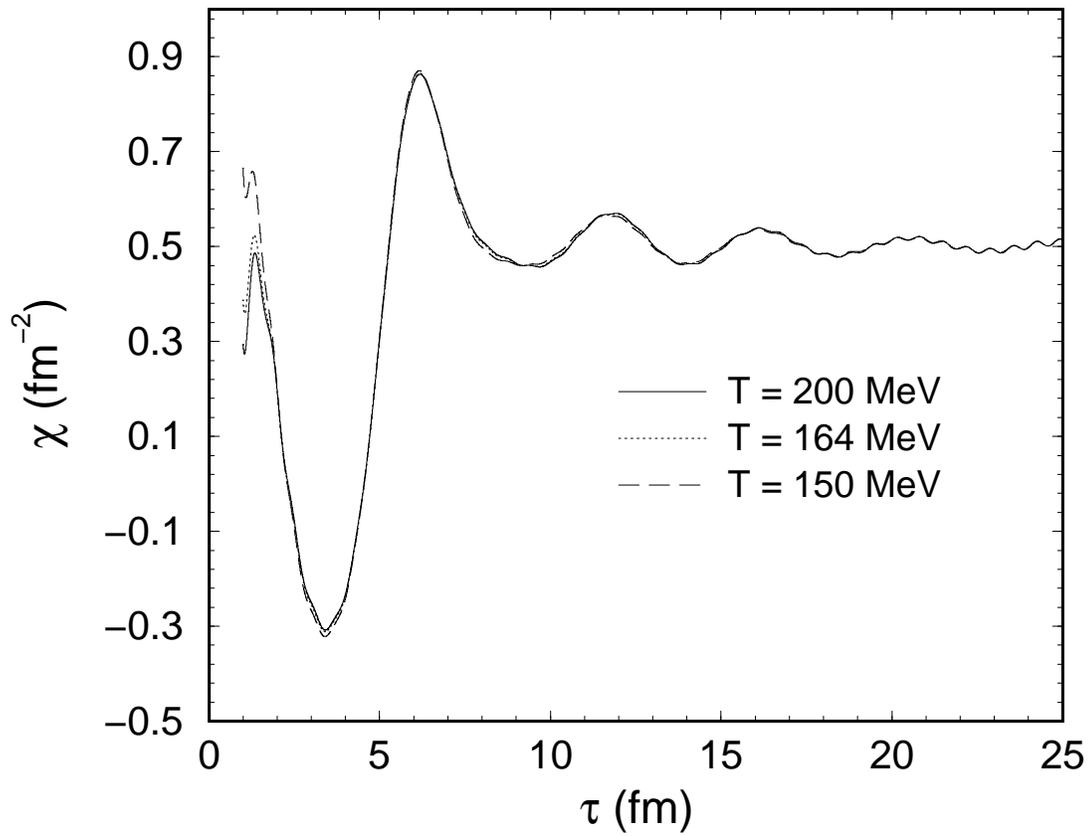}}
\caption{Proper time evolution of the $\chi$ field for three different
initial thermal distributions with $T$ = 200, 164, 150 MeV for the
initial conditions $\sigma(\tau_0) = \sigma_{T}$, $\pi_i(\tau_0) = 0$,
and $\dot\sigma(\tau_0) = -1$.}
\label{fig:temp}
\end{figure}

\begin{figure}
\epsfxsize = 6.0in
\centerline{\epsfbox{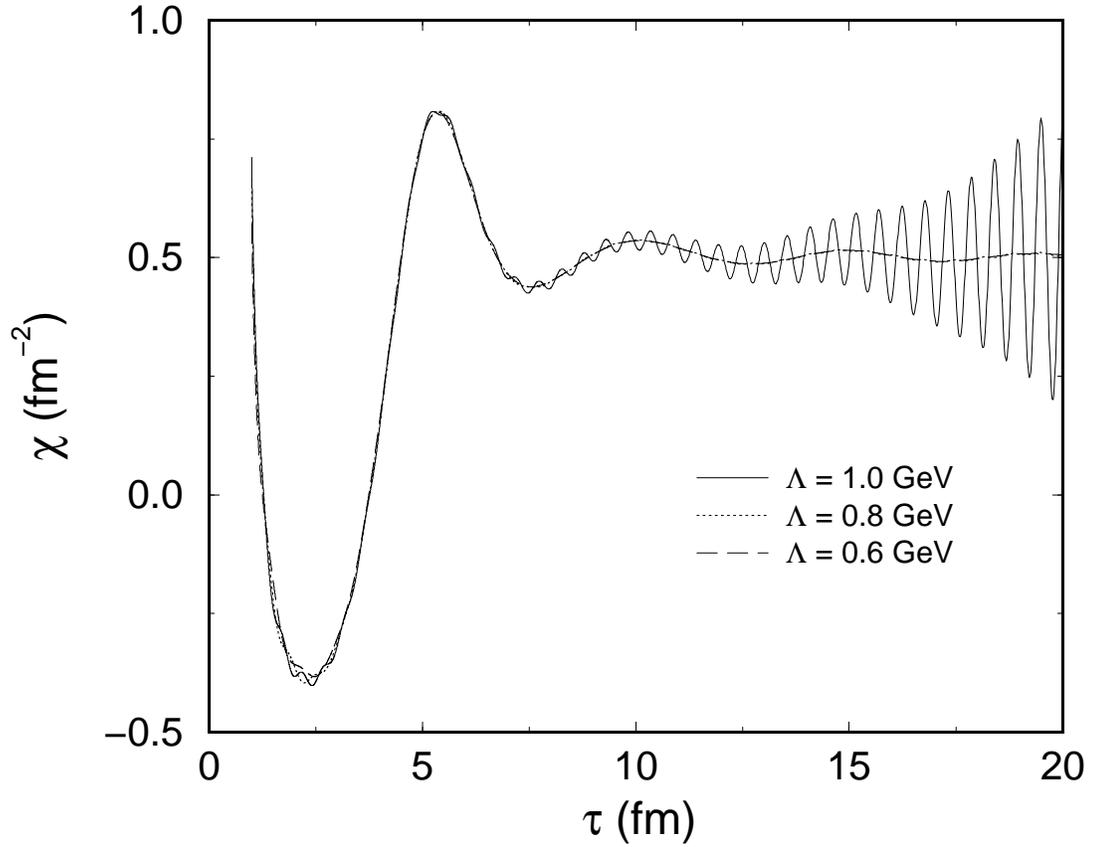}}
\caption{Proper time evolution of the $\chi$ field for three different
values of the cutoff $\Lambda$, with $\Lambda = 600, 800, 1000$ MeV
for the initial conditions $\sigma(\tau_0) = \sigma_{T}$,
$\pi_i(\tau_0) = 0$, and $\dot\sigma(\tau_0) = -1$.}
\label{fig:Lambda}
\end{figure}

In Fig.~\ref{fig:temp}, we show the effect of the initial
temperature on the evolution of the auxiliary field. We see that
varying the initial temperature has little effect. In
Fig.~\ref{fig:Lambda}, we show the evolution of the $\chi$ field for
different values of the cutoff $\Lambda$. We can see that $\chi$ is
independent of $\Lambda$, which shows that the renormalization has
been carried out correctly. In our simulations we use the value
$\Lambda =$ 800 MeV, since $\Lambda =$ 1 GeV is too close to the
Landau pole (see\cite{ref:cooperdcc}). When one chooses a cutoff too
close to the Landau pole the late time behavior becomes unstable. 

\begin{figure}
\epsfxsize = 6.0in
\centerline{\epsfbox{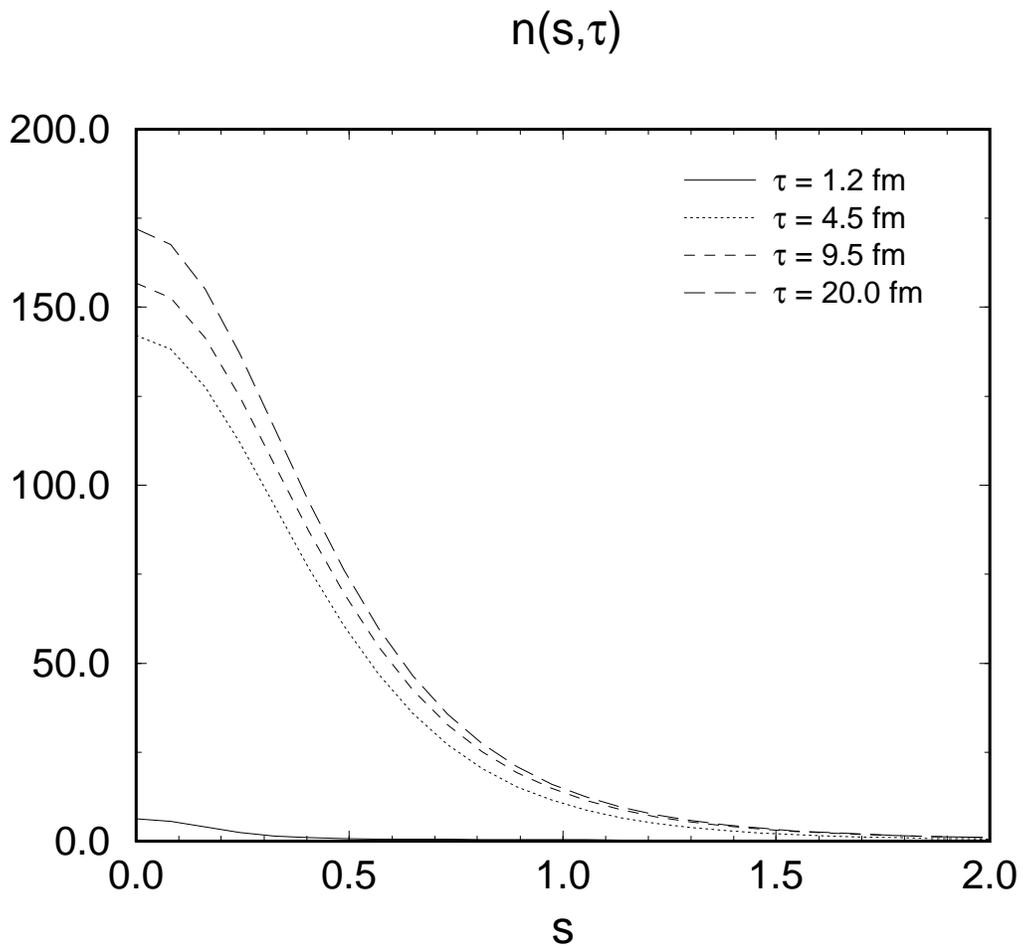}}
\caption{$n_s(\tau)$ computed from (\ref{nstau}), for the initial
conditions $\sigma(\tau_0) = \sigma_{T}$, $\pi_i(\tau_0) = 0$,
and $\dot\sigma(\tau_0) = -1$.}
\label{fig:ns1}
\end{figure}

\begin{figure}
\epsfxsize = 6.0in
\centerline{\epsfbox{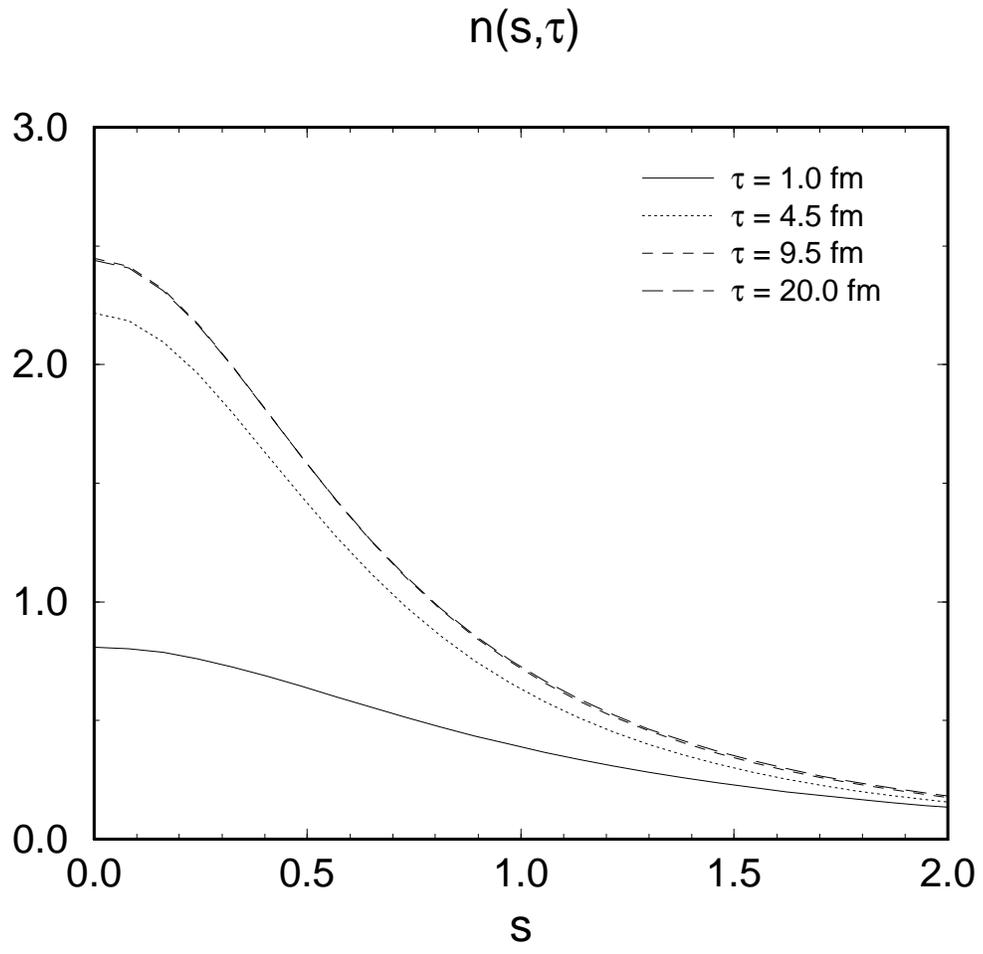}}
\caption{Same as the previous figure, but for the initial conditions
$\sigma(\tau_0) = \sigma_{T}$, $\pi_i(\tau_0) = 0$, and
$\dot\sigma(\tau_0) = 0$.}
\label{fig:ns0}
\end{figure}

Figures \ref{fig:ns1} and \ref{fig:ns0} show the number density
calculated from (\ref{nstau}), for several different proper
times. Figure \ref{fig:ns1} is a case where instabilities have arisen in
the system, and there is a large amount of particle production during
the time that $\chi$ has gone negative. Figure \ref{fig:ns0} is a case
with no instabilities, and there is very little particle production as
a function of proper time.

\begin{figure}
\epsfxsize = 6.0in
\centerline{\epsfbox{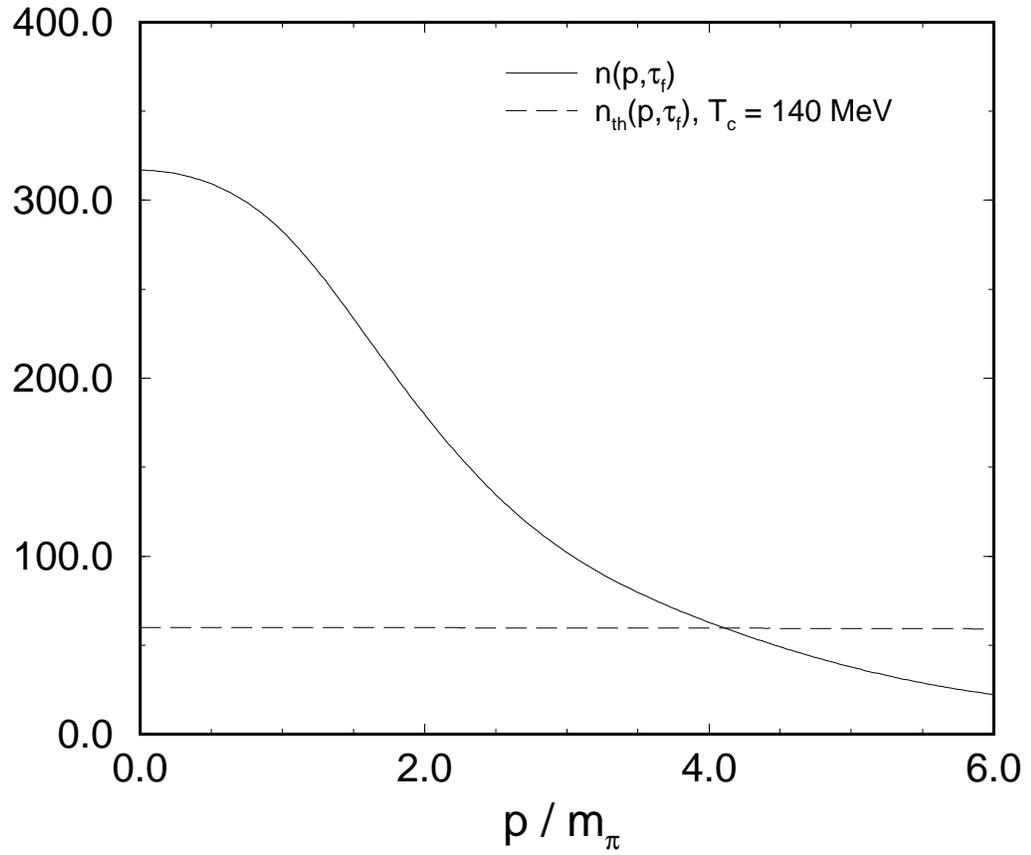}}
\caption{$n(p,\tau)$ computed from (\ref{eq:c-f2}) and 
$n_{th}(p, \tau)$ computed from (\ref{eq:c-f3}), for the 
same initial conditions as Fig.~\ref{fig:ns1}.}
\label{fig:np1}
\end{figure}

Figures \ref{fig:np1} and \ref{fig:np0} show the same distributions
transformed to the physical momentum $p$, as discussed in Appendix
\ref{app:nphys}. The momentum $p$ is plotted in units of $m_\pi$. We
compare these distributions to a hydrodynamical model calculation [see
(\ref{eq:c-f3})], where we have assumed that when the system reaches
the ``out'' regime, the final distribution is a combination of a
thermal distribution in the comoving frame at $T_c = m_\pi$ boosted to the
center of mass frame using the boost variable $\eta(r,t)$ (see
\cite{ref:cooperlandau}). For comparison purposes, we have
renormalized the thermal distributions to give the same center of mass
energy ($E = 100$ GeV) as the corresponding non-thermal distributions.
We see that as a result of the non-equilibrium evolution, there is an
enhancement at low momentum independent of whether or not there are
instabilities; however, the effect of instabilities is to greatly
magnify this low momentum enhancement.

\section{Conclusions}

In this paper we have performed numerical simulations to examine the
chiral phase transition during a uniform spherical expansion of the
hadronic plasma. We used the linear $\sigma$ model to leading order 
in a large-$N$
expansion and studied a wide range of initial conditions starting
above the critical temperature for the phase transition. We found
initial conditions which drove the system to instabilities that lead
to the formation of disoriented chiral condensates. Due to the
necessity of a rather large renormalized coupling constant, the
formation of instabilities lasted only a short time because of the 
large exponentially growing quantum corrections in (\ref{eq:chiu}), 
and no significant amount of domain formation was observed. However, 
we find that the phase space number density for our non-equilibrium 
evolution is significantly different from one which would result from 
an evolution in thermal equilibrium. The experimental signature for 
domain formation is an increase in the pion particle production rate 
at low momentum.  Our calculations were done in a mean-field 
approximation, where all the mode coupling is due to the presence of 
this mean field. In next order in large-$N$, scattering in the background 
mean field occurs, and the possibility for re-equilibrization exists. 
These effects will be incorporated in a future calculation.  
In comparison to particle production during a longitudinal expansion, 
we find that in a spherical expansion the system reaches the ``out'' 
regime much faster and more particles get produced. However the size 
of the unstable region, which is related to the size of the domain 
of DCCs, is not enhanced.

\section*{acknowledgements}

The authors would like to thank Emil Mottola, Yuval Kluger and Salman
Habib for useful discussions. UNH gratefully acknowledges support by
the U.S. Department of Energy (DE-FG02-88ER40410). One of us (FC)
would like to thank the University of New Hampshire for its
hospitality. 


\appendix
\section{Properties of the $\pi_{sl}$ Functions}

\label{app:piprop}

In this appendix, we discuss properties of the functions 
$\pi_{sl}(\eta)$, which are {\em real} solutions of the equation,
\begin{displaymath}
   \frac{1}{ \sinh^2 \eta } 
      \frac{\partial}{\partial \eta} \left( \sinh^2 \eta \,
      \frac{\partial\pi_{sl}}{\partial \eta} \right) +  
      \left\{ s^2 + 1 - \frac{ l(l+1) }{ \sinh^2 \eta } 
      \right\} \pi_{sl}
    = 0   \>,
\end{displaymath}
or,
\begin{displaymath}
   \frac{\partial^2 \pi_{sl}}{\partial \eta^2 } + \frac{2}{\tanh \eta}
      \frac{\partial\pi_{sl}}{\partial \eta} + 
      \left\{ s^2 + 1 - \frac{ l(l+1) }{ \sinh^2 \eta } 
      \right\} \pi_{sl}
    = 0   \>,
\end{displaymath}
for $\eta$ in the range: $0 \leq \eta \leq \infty$. 
With the substitution,
\begin{displaymath}
   \pi_{sl}(\eta) = u_{sl}(\eta) / \sinh \eta
      \>,
\end{displaymath}
we find that $u_{sl}(\eta)$ satisfies:
\begin{displaymath}
   u_{sl}'' + \left[ s^2 - \frac{ l(l+1) }{ \sinh^2 \eta } \right]
      u_{sl} = 0  \>.
\end{displaymath}
The general solution is given by\cite{ref:BirDav}:
\begin{displaymath}
   \pi_{sl}(\eta) = \frac{\sinh^l \eta}{M_{sl}} 
      \left(\frac{d}{d\cosh\eta}\right)^{(1+l)} \cos(s\eta)
\>,
\end{displaymath}
where the normalization $M_{sl}$ is given by
\begin{displaymath}
   M_{sl} = \sqrt{(\pi/2) s^2 ( s^2 + 1^2 ) \cdots (s^2 + l^2)}
\>.
\end{displaymath}
The completeness relation is given by:
\begin{displaymath}
   \int_0^\infty ds \,
      \pi_{sl}(\eta) \, \pi_{sl}(\eta') 
      = \delta(\eta - \eta') / [ \sinh \eta \sinh \eta' ] \>.
\end{displaymath}
For the functions ${\cal Y}_{slm}$ defined in (\ref{eq:bigy}), the addition
formula is\cite{ref:parker}:
\begin{eqnarray*}
   && \sum_{lm} {\cal Y}_{slm}^{\ast}(\eta_1,\theta_1,\phi_1) \,
   {\cal Y}_{slm}(\eta_2,\theta_2,\phi_2)
   \\
   && \phantom{\sum_{lm}}
   =  \frac{s}{ 2 \pi^2 } \, 
      \frac{ \sin (s \eta) }{ \sinh \eta }
   = \frac{ s^2 }{ 2 \pi^2 } \, 
   \left\{ 1 - \frac{ s^2 + 1 }{ 6 } \, \eta^2 + \ldots \> \right\}
\>,
\end{eqnarray*}
where $\eta$ is defined by:
\begin{eqnarray*}
   \cosh \eta & = & \cosh \eta_1 \, \cosh \eta_2 
      - \sinh \eta_1 \, \sinh \eta_2 \, \cos \theta
   \\
   \cos \theta & = & 
      \cos \theta_1 \, \cos \theta_2 + 
      \sin \theta_1 \, \sin \theta_2 \, \cos( \phi_1 - \phi_2 ) 
   \>.
\end{eqnarray*}
Therefore, taking the limit, 
$(\eta_1,\theta_1,\phi_1) \rightarrow (\eta_2,\theta_2,\phi_2)$,
or $\eta \rightarrow 0$, we find
\begin{displaymath}
   \sum_{lm} | {\cal Y}_{slm}(\eta,\theta,\phi) |^2 = 
   \frac{ s^2 }{ 2 \pi^2 }  \>.
\end{displaymath}
By differentiating both sides of the addition formula, 
we can show that
\begin{eqnarray*}
   \sum_{lm} \left| 
      \frac { \partial {\cal Y}_{slm}(\eta,\theta,\phi) }
            { \partial \eta } \right|^2 
   & = &  \frac{ s^2 }{ 2 \pi^2 } 
      \left(  \frac{ s^2 + 1 }{ 3 } \right) 
      \>,  \\
   \sum_{lm} \left| 
      \frac { \partial {\cal Y}_{slm}(\eta,\theta,\phi) }
            { \partial \theta } \right|^2 
   & = &  \frac{ s^2 }{ 2 \pi^2 } 
      \left(  \frac{ s^2 + 1 }{ 3 } \right) \,
      \sinh^2 \eta  
      \>,  \\
   \sum_{lm} \left| 
      \frac { \partial {\cal Y}_{slm}(\eta,\theta,\phi) }
            { \partial \phi } \right|^2 
   & = &  \frac{ s^2 }{ 2 \pi^2 } 
      \left(  \frac{ s^2 + 1 }{ 3 } \right) \,
      \sinh^2 \eta \, \sin^2 \theta  
      \>. 
\end{eqnarray*}

\section{Transformation to Physical Variables}
\label{app:nphys}

In terms of the initial distribution of particles $n_s(\tau_0)$ and
$\beta$ we have:
\begin{displaymath}
   n_s(u) = n_s(u_0) + |\beta(s,u)|^2 ( 1+2n_s(u_0))
\>,
\end{displaymath}
where $n_s(u)$ is the adiabatic invariant interpolating phase space
number density which becomes the actual particle phase space number
density in the comoving frame when interactions have ceased. We now
need to relate this quantity to the physical spectra of particles
measured in the lab. At late $\tau \ge \tau_f \approx 10$ fm our system 
relaxes to the vacuum and $\chi$ becomes the square of the physical pion mass
$m^2$. The comoving center of mass energy of outgoing particles can
then be identified with
\begin{displaymath}
   \omega_s (\tau_f) = \sqrt{ {s^2 \over \tau_f^2}+ m^2}
\>.
\end{displaymath}

The actual distribution of momenta in the lab frame is a combination
of the collective (``fluid'') motion described by the boost $\eta$
from the comoving frame to center of mass frame and the comoving
particle distribution. Here, the space-like hypersurface on which one
is counting particles is at fixed proper time $\tau_f$. This
distribution is given by the Cooper-Frye formula\cite{ref:c-f} which is:

\begin{equation}
   E {dN \over d^3 p} = E {dN \over 4 \pi p^2 dp} 
      = \int f(x, p) p^{\mu} d \sigma_{\mu}
\label{eq:c-f}
\>.
\end{equation}

We identify the relativistic phase space distribution function $f(x,p)$
with $ n_s(u_f)$. The dependence of $s$ on the space time variable $x$
and the outgoing momentum $p$ is found from the relationship 
\begin{displaymath}
   p^{\mu} u_{\mu} = \omega_s (\tau_f) 
      = \sqrt{ {s^2 \over \tau_f^2} + m^2}
\>.
\end{displaymath}
We choose the measured momentum $p$ to be in the $z$ direction
$\epsilon_3$ of our spherical coordinate system. We have that
\begin{eqnarray*}
   u^{\mu} &=& (\cosh \eta,  \sinh \eta ~ {\hat \epsilon}_r)
   \nonumber \\
   p^{\mu} &=& ( E, p ~{\hat\epsilon}_3 )
\>,
\end{eqnarray*}
so that
\begin{displaymath}
   p^{\mu} u_{\mu} =  E \cosh \eta - p \cos \theta \sinh \eta
\>.
\end{displaymath}
 
The surface on which one is counting particles is the time-like surface
$\tau = \tau_f$  with
\begin{displaymath}
   d \sigma_{\mu} = \left(1, - {\partial t_f \over \partial r} 
   {\hat \epsilon}_r \right) d^3 r
\>.
\end{displaymath}
Changing variables from $r$ to $\eta$ at fixed $\tau$ we then obtain
\begin{eqnarray}
   E {dN \over 4 \pi p^2 dp} &=& n(p, \tau)
   \nonumber \\
   &=& \int f(x, p) d\eta~ d\cos \theta ~\tau_f^3
      ~\sinh^2 \eta ~p^{\mu} u_{\mu}
\>,
\label{eq:c-f2}
\end{eqnarray}
where
\begin{displaymath}
   p^{\mu} u_{\mu} =  E \cosh \eta - p \cos \theta \sinh \eta
\end{displaymath}
and we have used the isotropy assumption and chosen $p$ as the $z$ axis.
Here $E = \sqrt{p^2 + m^2}$.  

The calculation that (\ref{eq:c-f2}) needs to be compared with is a
hydrodynamical model calculation for a local thermal equilibrium
flow.  In a hydrodynamical model of heavy ion
collisions\cite{ref:bjorkenlandau,ref:cooperlandau}, the final
spectrum of pions is given by a combination of the fluid flow and a
local thermal equilibrium distribution in the comoving frame. One
calculates this spectrum at the critical temperature $T_c(x,t)$ when
the energy density goes below
\begin{displaymath}
   \epsilon_c = {1 \over (\hbar/mc)^3}
\>.
\end{displaymath}
This defines the breakup surface $\tau_c$ , after which the particles
no longer interact so that this distibution is frozen at that
temperature. For an ultrarelativistic gas of pions, this occurs when
$T_c= m$.  The covariant form for the spectra of particles is again
given by the Cooper-Frye formula \cite{ref:c-f}:
\begin{equation}
   E {dN \over 4 \pi p^2 dp} = \int f(x, p) d\eta~ d\cos \theta ~\tau_c^3
      ~\sinh^2 \eta ~p^{\mu} u_{\mu}
\label{eq:c-f3}
\end{equation}
but now $f(x,p)$ is the single particle relativistic phase space
distribution function for pions in local thermal equilibrium at a
comoving temperature $T_c(\tau_c)$:
\begin{displaymath}
   f(x,p) =\{ {\rm exp} [p^{\mu}u_{\mu}/T_c] -1 \}^{-1}
\>. 
\end{displaymath}
We have identified the left hand side of (\ref{eq:c-f3}) 
as $n_{th}(p,\tau)$.



\begin{figure}
\epsfxsize = 6.0in
\centerline{\epsfbox{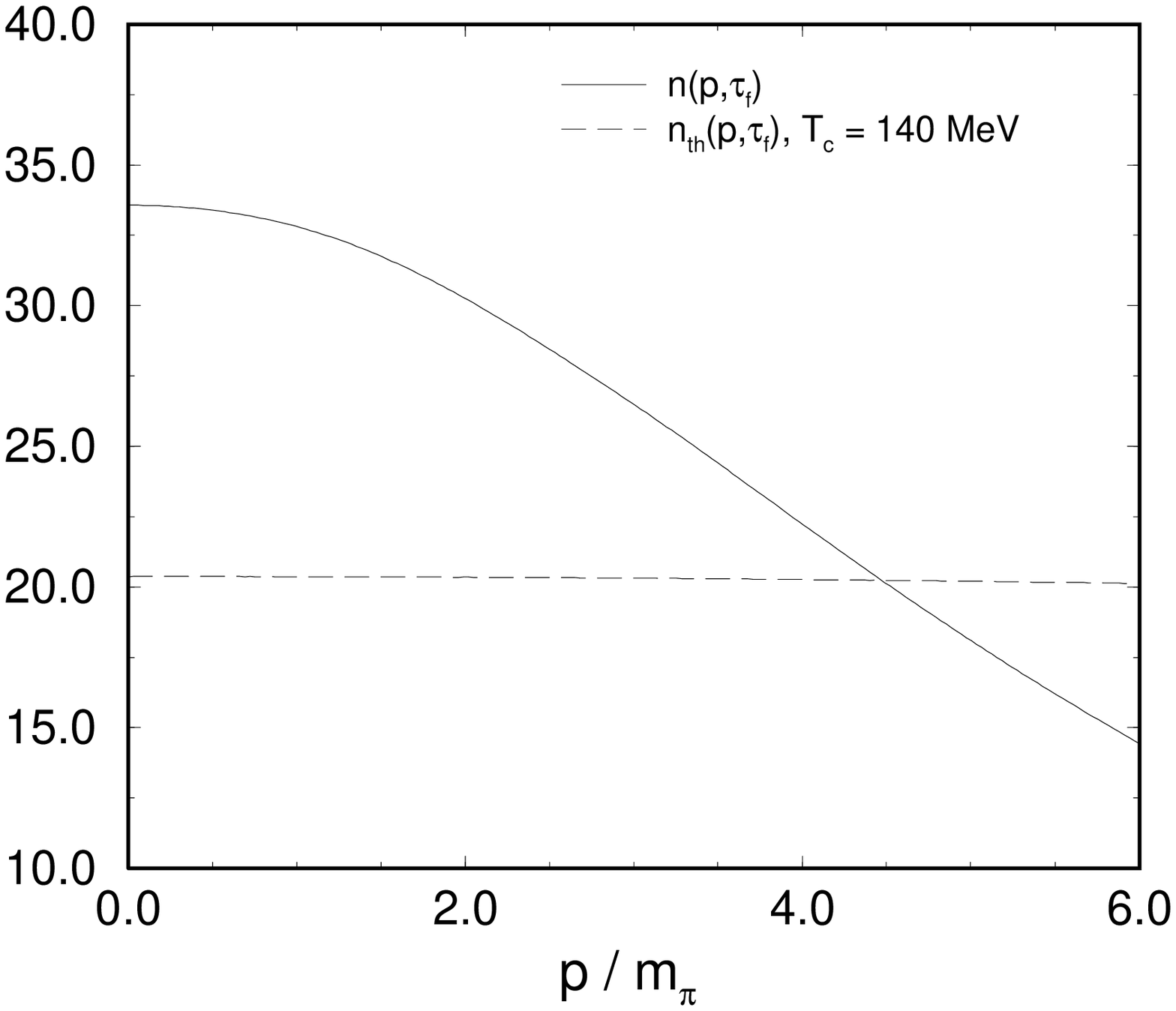}}
\caption{Same as the previous figure, but
for the same initial conditions as Fig.~\ref{fig:ns0}.}
\label{fig:np0}
\end{figure}



\begin{references}

\bibitem{ref:anselm}
A. Anselm, \pl B {\bf 217}, 169 (1989); A. Anselm and
M. Ryskin, {\em ibid} {\bf 226}, 482 (1991)

\bibitem{ref:bjorken}
J.D. Bjorken, Int.\ J.\ Mod.\ Phys.\ A {\bf 7}, 4189 (1992); Acta\
Phys.\ Pol.\ B {\bf 23}, 561 (1992).

\bibitem{ref:rajwil}
K. Rajagopal and F. Wilczek, Nucl.\ Phys.\ {\bf B399}, 395 (1993);\\
S. Gavin, A. Gocksch and R.D. Pisarski, \prl {\bf 72}, 2143 (1994);\\
S. Gavin and B. M\"uller, \pl B {\bf 329}, 486 (1994);\\
S. Gavin, hep-ph/9407368;\\
D. Boyanovsky, H.J. de Vega, and R. Holman, \prd {\bf 51}, 734 (1995);\\
J.-P. Blaizot and A. Krzywicki, \prd {\bf 46}, 246 (1992);
\prd {\bf 50}, 442 (1994).

\bibitem{ref:cent} 
C.M.G. Lattes, Y. Fujimoto, and S. Hasegawa,
Phys.\ Rep.\ {\bf 65}, 151 (1980).

\bibitem{ref:cooperdcc}
F.~Cooper, Y.~Kluger, E.~Mottola, and J.~P.~Paz, \prd {\bf 51}, 
2377 (1995).

\bibitem{ref:bjorkenlandau}
J.~D.~Bjorken, \prd {\bf 27}, 140 (1983).

\bibitem{ref:cooperlandau}
Fred Cooper, Graham Frye, and Edmond Schonberg, \prd 
{\bf 11}, 192 (1975).

\bibitem{ref:Stanley}
H.~E.~Stanley, Phys.\ Rev.\ {\bf 176}, 718 (1968); \\
K.~Wilson, \prd {\bf 7}, 2911 (1973); \\
S.~Coleman, R.~Jackiw, and H.~D.~Politzer, \prd {\bf 10},
2491 (1974).

\bibitem{ref:ctp}
   J.~Schwinger, 
   J.\ Math.\ Phys.\ {\bf 2}, 407 (1961);\\
   P.~M.~Bakshi and K.~T.~Mahanthappa,
   J.\ Math.\ Phys.\ {\bf 4}, 1 (1963); {\bf 4}, 12 (1963);\\
   L.~V.~Keldysh,
   Zh.\ Eksp.\ Teo.\ Fiz.\ {\bf 47}, 1515 (1964)\  
   [Sov.\ Phys.\ JETP {\bf 20}, 1018 (1965)];\\
   G.~Zhou, Z.~Su, B.~Hao and L.~Yu, 
   Phys.\ Rep.\ {\bf 118}, 1 (1985);\\
   F. Cooper, S. Habib, Y. Kluger, E. Mottola, J.P. Paz and Paul
   Anderson, \prd {\bf 50}, 2848 (1994).

\bibitem{ref:parker}
L. Parker and S. A. Fulling, \prd {\bf 9}, 341 (1974).

\bibitem{ref:cooperren}
F. Cooper and E. Mottola, \prd {\bf 36}, 3114 (1987).

\bibitem{ref:mottola}
Salman Habib, Yuval Kluger, Emil Mottola, and Juan Pablo Paz, Los Alamos
Preprint (LA-UR-95-3393/LBL-37786), 1995.

\bibitem{ref:callan}
C. G. Callan, S. Coleman, and R. Jackiw, Ann.\ Phys.\ (NY) 
{\bf 59}, 42 (1970).

\bibitem{ref:BirDav}
N.D. Birrell and P.C.W. Davies, {\em Quantum Fields in Curved Space}, 
(Cambridge University Press, 1982), p. 122.

\bibitem{ref:c-f} F. Cooper and G. Frye, \prd {\bf 10}, 186 (1974).
See also ``Fireball Spectra'' by Ekkard Schnedermann, Josef Sollfrank,
and Ulrich Heinz in {\em Particle Production in Highly Excited
Matter}, pp. 175-206, edited by H.H. Gutbrod (Plenum, New York, 1993).

\end{references}
\end{document}